\documentclass[aps,twocolumn]{revtex4}
\usepackage{graphicx,subfigure}
\usepackage{amsmath,amssymb}
\usepackage{bm}

\newcommand{\be}{\begin{eqnarray}}
\newcommand{\ee}{\end{eqnarray}}

\def\slashchar#1{\setbox0=\hbox{$#1$}           % set a box for #1 
   \dimen0=\wd0                                 % and get its size
   \setbox1=\hbox{/} \dimen1=\wd1               % get size of /
  \ifdim\dimen0>\dimen1                        % #1 is bigger
 \rlap{\hbox to \dimen0{\hfil/\hfil}}      % so center / in box
  #1                                        % and print #1
 \else                                        % / is bigger
    \rlap{\hbox to \dimen1{\hfil$#1$\hfil}}   % so center #1
    /                                         % and print /
 \fi}                                         %

\begin{document}

\title{Instanton-dyon Ensembles III: Exotic Quark Flavors}

\author{ Rasmus Larsen   and Edward  Shuryak }

\affiliation{Department of Physics and Astronomy, Stony Brook University,
Stony Brook NY 11794-3800, USA}

\begin{abstract}
``Exotic quarks" in the title refers to a modification of quark periodicity condition on the thermal circle by introduction
of some phases
 -- known also as ``flavor holonomies" --  different quark flavors.
These phases provide a valuable tool, to be used for better understanding of deconfinement and chiral restoration phase transitions: by changing them one can dramatically modify both phase transitions. In the language of  
instanton constituents -- instanton-dyons or monopoles -- it has a very direct explanation: 
the interplay of flavor and color holonomies
can switch topological zero modes between various dyon types.  
The model we will study in detail, the so called $Z_{N_c}$-symmetric QCD model with
equal number of colors and flavors $N_c=N_f=2$ and special arrangement of flavor and color holonomies,
 ensure ``most democratic" setting, in which each quark flavor and each dyon type are in one-to-one
 correspondence. The usual QCD has the opposite ``most exclusive" arrangement:  all quarks are antiperiodic
 and thus all zero modes fall on only one -- twisted or $L$ -- dyon type.   
As we show by ensemble simulation,   deconfinement and chiral restoration phase transitions in these two models
are dramatically different. In the usual QCD both are smooth crossovers: but in the case of  $Z_2$-symmetric
model deconfinement becomes strong first order transition, while   chiral symmetry remains broken for all
 dyon densities studied. These results are in good correspondence with those from recent lattice simulations.

\end{abstract}
\maketitle

\section{Introduction}

QCD-like gauge theories display two main non-perturbative phenomena,  {\em confinement} and {\em spontaneous 
breaking of $SU(N_f)$ chiral symmetry.}  
Their mechanism has been discussed extensively since the 1970's:
We will mention some historical highlights below. 
Our previous papers, to be referred to as I \cite{Larsen:2015vaa} and II   \cite{Larsen:2015tso}
 below, addressed those phenomena in a model based on a certain gauge topology model.

Before going to details of that model, let us first explain the main new element of this paper,
what do we mean by {\em exotic quarks}, and why one should be interested in such theories.
Standard Euclidean formulation of QCD-like theories define time $\tau\in S^1$, on the so called Matsubara circle 
with a circumference $\beta=\hbar/T$.  Bosons -- the gauge fields -- are $periodic$ on this circle, and
fermions  are  $anti-periodic$: as a result standard Bose and Fermi thermal factors 
automatically appear in thermodynamical expressions. One option to make
quarks {\em exotic } is to generalize their periodicity condition on  $S^1$ by adding some
arbitrary phase to it, and this is the one to be used in this work. 
With a generic value of this phase, quarks are neither fermions nor bosons:
perhaps one can use a term originated from condensed matter physics -- the ``anyons".

 Short answer to why  such studies can be instructive is as follows:  Chiral symmetry breaking is associated with
Dirac zero modes of certain topological solitons. By changing the quark phases, one can very
effectively manipulate their coupling to solitons of different kinds, and thus learn which of them are in fact relevant
for the phenomena under consideration.

Since this is our third paper of the series, it hardly needs an extensive introduction.
Here is a brief outline of the historic development of  non-perturbative QCD. 
Since the 1970's, $confinement$ was historically connected with gauge field solitons with magnetic charge -- the monopoles.
It is sufficient to mention the so called ``dual superconductor" model by 't Hooft and Mandelstam \cite{dualsuper} 
and Polyakov's prove of confinement in 2+1-dimensional gauge theories \cite{Polyakov:1976fu}.  

{\em Chiral symmetry breaking} has been first addressed by Nambu and Jona Lasinio \cite{NJL} back in 1961,
who pointed out even the existence of the chiral symmetry, as well as its spontaneous breaking by a
nonzero quark condensate. It was shown that in order to create it, some strong attractive 4-fermion interaction is needed, at momenta below
a cutoff $Q< \Lambda \sim 1\, GeV$. Twenty years later the emerging instanton phenomenology 
 \cite{Shuryak:1981ff} made it clear that this interaction is nothing else but
 multi-fermion effective forces induced by instantons. Furthermore, unlike the effective Lagrangian conjectured by NJL,
 it also explicitly breaks $U(1)_a$ axial symmetry. 
With time somewhat different perspective  on chiral symmetry breaking had developed, focused on Dirac eigenvalue spectrum. 
The low lying ones are formed via  collectivization  of the
topological fermionic zero modes  into the so called 
{\em Zero Mode Zone} (ZMZ): for a review see \cite{Schafer:1996wv} .

While the main non-perturbative phenomena were for a long time ascribed to two  rather distinct  mechanisms, 
the results of the numerical simulations on the lattice 
 persistently indicated that  the  
deconfinement temperature $T_{deconf}$ and chiral symmetry restoration, one  to be called $T_\chi$,
are in fact very close, at least for near-real-world QCD (with the number of colors and flavors $N_c=3,N_f=2+1$
and quark masses small). This fact, and in general  their 
 interplay  in all QCD-like theories became one of the most puzzling issues
in non-perturbative QFT.

But does it really hold for all   QCD-like theories? One simple way 
 to modify QCD is to change the color charge (representation) of the quark fields.
Changing quark {\em color charge} from the fundamental to adjoint representation, Kogut \cite{Kogut:1986jt}
  observed in $SU(2)$ gauge theories (with 2 and 4 species of Majorana fermions, 
  or 1 and 2 species of the Dirac ones)   
 two distinct phase transitions, at  temperatures  separated by a large factor
  \be {T_\chi^{aQCD} \over  T_{deconf}^{aQCD} }\sim O(10)  \ee 
Later studies for the $SU(3)$ theories with adjoint quarks  
 \cite{Karsch:1998qj,Cossu:2008wh} also observed such a large separation.
 Such significant split between two transition temperatures may suggest that 
 some quite different mechanisms are responsible for these two phenomena.
Our studies of the adjoint quarks will be described  in
a separate publication  IV \cite{LSadj}.

On the other hand,  recent theory developments had revealed some
common dynamical roots of  confinement and chiral symmetry breaking. When in 1998 
the finite-temperature instantons were generalized to 
 the case of non-zero expectation value of the Polyakov loop, by  Lee-Li-Kraan-van Baal \cite{Kraan:1998sn,Lee:1998bb}, it had been realized that  each consist of $N_c$ objects, known as instanton-dyons (or instanton-monopoles).
Unlike instantons themselves, their constituents have charges and directly back-react on the holonomy
value. This observation lead to the proposal \cite{Diakonov:2009ln} of confinement being driven by this effect, 
and in a very specialized setting it has been shown to induce confinement \cite{Poppitz:2011wy,Poppitz:2012sw},
even for exponentially small dyon density. Recent works using mean field methods \cite{Liu:2015ufa,Liu:2015jsa}
and our own two papers \cite{Larsen:2015vaa,Larsen:2015tso} had shown that sufficiently dense ensemble of the dyons 
does generate both confinement and chiral symmetry breaking, with close transition temperatures (too close to
separate inside the errors).

Let us now turn to discussion of quarks with non-standard   periodicity phases
 on $S^1$: can those have any measurable physical effect on observables? 
 Already
first experimentation with quark phases  \cite{Gattringer:2001yu} answered this question affirmatively. Even with fixed
quenched lattice gauge configurations, it was confirmed 
that the value of the quark condensate depends on the 
 periodicity phase. 
In particular, the ``periodic" (or bosonic)  quarks possesses chiral symmetry breaking till very high $T$,
in striking contrast to the usual antiperiodic ones. Suggested explanation \cite{Shuryak:2012aa}
was formulated in terms of the instanton-dyons: 
the periodic quarks has zero modes with  M-type dyons, while
the antiperiodic quarks have zero modes
with ``twisted" L-type dyons. The ``masses" (actions) of those  are different at $T>T_{conf}$, the twisted ones $L$
are heavier $S_L > S_M$ and therefore have smaller density
\be n_M \gg n_L \ee  
So, the  periodic quarks ``see" much denser ($M-$dyon) ensemble than the antiperiodic ones, which explains
 larger quark condensate and higher $T\chi$.

In a framework of PNJL model, 
%Further development of these ideas has been carried out by 
Kouno et al \cite{K1,K2,K3,K4,K5} suggested to require different boundary conditions to
different quark flavors. For enhanced symmetry they focus on theories in which the
number of colors and flavors are the same, $N_c=N_f$.
In particularly, for the SU(3) color, they use quark angles for $u,d,s$
quarks to be of the form $(0,\theta,-\theta)$, respectively. The parameter $\theta$ can be varied from
 $\theta=0$ (the usual QCD) to $\theta=2\pi/3$, at which point the theory becomes ``center symmetric".
Thus the name of the resulting model, ``$Z_{N_c}$-symmetric QCD". In the framework
of the PNJL model, these authors found substantial dependence of $T_\chi$ on $\theta$. The effect is the largest for the
symmetric value
\be T_\chi^{Z3QCD}  \approx 2 T_{deconf}^{Z3QCD}   \ee 
Of course, PNJL is just a model, using as input, the holonomy potential and the 4-quark NJL-like lagrangian,
the same as for ordinary QCD, with $unmodified$ parameters: whether this can be justified is unclear.

The flavor-dependent phases had also been suggested in the framework of supersymmetric QCD
in \cite{Poppitz:2013zqa}.

Misumi et al \cite{Misumi:2015hfa} had recently put this $Z(3)$-symmetric theory on the lattice.  They observe 
that compared to ordinary QCD, in this case the deconfinement transition is significantly
strengthened to  the first order phase transition, with clear hysteresis etc. 
The chiral breaking for $Z(3)$-symmetric theory
is indeed present at higher  temperatures. It is hard to tell from the paper, since small mass
extrapolation is not yet performed, whether the chiral symmetry is in fact restored at any temperatures. 
 Furthermore, the
chiral condensates for different flavors become clearly different, so $T_\chi$ should get split for different flavors.
Qualitatively, the PNJL-based  conclusions are confirmed.

\section{$Z_{N_c}$-symmetric QCD and the instanton-dyons }

Let us start this section by reminding that in the framework of the original instanton model
most of the phenomena mentioned in the Introduction would be impossible to explain. Indeed,
the number of zero modes of the instanton is prescribed by the topological
index theorem and is independent on the periodicity condition. 

On the other hand, after it has been recognized that instantons has to be split into instanton-dyons, the situations changes dramatically. Indeed, quarks with different boundary angles 
can be coupled to different types of dyons. Dialing different values of those angles, one can
see the consequences from which it will  eventually be
possible to understand the dynamical role of those objects. 

The ``$Z_{N_c}$-symmetric QCD" proposed by Kouna et al does indeed have outstandingly
simple symmetry properties in the  instanton-dyon model: {\em each quark flavor has zero modes with
a different type of   instanton-dyons}. This means that each quark flavor has its own ``dyon plasma"
with which it interacts.
We remind that in this model the number of colors and flavors must match, $N_f=N_c$, so the number of quark and dyon types match as well.

Furthermore, in the low $T$, near and below $T_{deconf}$ the holonomy
values tends to the symmetric ``confining" value, at which all types of the dyons obtain the same action.
This fact indeed made the model $Z_{N_c}$-symmetric. 

In the opposite limit of high $T$, the holonomy moves to trivial value, and the actions of different dyons
become distinct. This implies that   each quark flavor has its own ``dyon plasma" with distinct densities,
leading to flavor-dependent $T_\chi$.
  
  One more qualitative idea
  is related to the values of
    $z$ which are ``intermediate" between the extremes discussed above.
    Those are values at which the zero modes jump from one kind of 
  the dyon to the next. This happens by ``delocalization" of the zero mode, which means that
  at such particular $z$ values the zero modes become long-range. Since in this case
  the ``hopping" matrix elements, describing quark-induced dyon-dyon interactions, get enhanced,
  one may also expect that the chiral condensate is effectively strengthened.  
  
\section{The setting of the simulations}

Let us remind the reader the setting used in our first paper \cite{Larsen:2015vaa} with instanton dyons.
Certain number of them -- 64 or 128 -- are placed on the 3-dimensional sphere. Its radius thus control the density.
Standard Metropolis algorithm is used to numerically simulate the distribution defined by 
classical and one-loop partition function. 
We studied the simplest non-Abelian theory with two colors $N_c=2$, which has a single holonomy parameter $\nu\in[0,1]$.
Free energy is calculated and the adjustable parameters of the model -- the value of the holonomy $\nu$ and the ratio of
densities of  $M,L$ type dyons -- are placed at its minimum. 

The partition function is $Z_2$ symmetric, under    $\nu \leftrightarrow \bar{\nu}=1-\nu$ and $M\leftrightarrow L$
replacement. A distinct {\em symmetric phase} has minimal free energy at the symmetric point $\nu=1/2$ ,
and equal number of $L,M$ dyons. {\em Asymmetric phase} has free energy with two minima, away from  
the center $\nu=1/2$: by default the
spontaneous breaking of  $Z_2$ is assumed to happen to smaller value of $\nu$, so that at high $T$ it goes
to zero. %It correspond to density of $L$ dyons smaller than that of $M$ ones.

%\section{The setting and the simulation parameters}

Following the paper \cite{Larsen:2015tso} we use the following parametrization of the overlap between zero-modes
\begin{eqnarray}
T_{ij} & =& v_k c'\exp{\left(-\sqrt{11.2+(v_kr/2)^2}\right)} ,\label{Tij}
\end{eqnarray}
where $v_k$ is $v=2\pi \nu$ for $M$ dyons, and $v_k$ is $\bar{v}=2\pi \bar{\nu} = 2\pi (1-\nu)$ for $L$ dyons.
The three constants in the model is the same as our previous paper and is: $x_0=2$ for the dimensionless size of the core. $\Lambda=4 $ for the overall constant and $-\log(c') = -2.6$ for the constant on $T_{ij}$.

The simulation have been done using standard Metropolis algorithm. An update of all $N=64$ or $128$ dyons corresponds to one cycle. Each run consist of 3000 cycles. Free energy is measured by standard trick, involving integration over the interaction parameter from zero to one.   The simulation was done on a $S^3$ circle, its volume is  $V=2\pi^2 r^3$:
we use $r$ in some places below.

The input ``action parameter" $S$ defines the instanton-dyon amplitude, and literally corresponds to
the sum of the $L$ and $M$ dyon actions in semiclassical amplitude. In one loop approximation
it is related to the temperature $T$ by the asymptotic freedom relation
\be S=({11N_c\over 3}-{2N_f \over 3}) log({T \over \Lambda_T}). \ee
In I, II we approximately related the constant $\Lambda_T$ to the phase transition temperature $T_c$:
we do not do it in this work because there is no single phase transition in the theory we study now.  

The varied parameters of the model include (1) The holonomy $\nu$ which is related to the Polyakov loop as $P=\cos(\pi \nu)$ and  (2,3) The densities  of $M$ and $L$ dyons $n_M,n_L$. 
After the free energy is defined for each run, the values of these
parameters, corresponding to its minimum, are fitted and used. 

Other parameters include
(4) The Debye mass, which is used to describe the falloff of the fields: its value is kept ``self consistent"
by a procedure explained in I. Finally we mention
 (5)  the auxiliary interaction variable  which is then integrated in order to obtain the free energy $F$.

The organization of the numerical sets were done as follows.  
 An initial survey found the areas of interest, corresponding to minima of the free energy and most important
 variations of the results. Then the final set of simulations has been performed: its parameters are summarized in the Table I.
In total 1170000 individual runs where done for the final set of data, from which the plots were made.

\begin{table}[h]
\begin{tabular}{ l | c | c | r  }
  \hline  
 &  Min & Max & Step size  \\   
   \hline 
$\lambda$ & 0 & 0.1 & 1/90  \\
$\lambda$ & 0.1 & 1.0 & 0.1  \\  
$\nu$ & 0.05 & 0.55 & 0.025  \\                    
$r$ & 1.2 & 1.8 & 0.05 \\
$N_M$ & 3 & 18 & 2  \\
$M_d$ & 1 & 3.5 & 0.5  \\
$S$ & 5 & 9.5 & 0.5  \\
  \hline  
\end{tabular}
\caption{The input parameters used for the final set of simulations. The step sizes given are some standard ones: yet
some areas was given extra attentions. For example  around $N_M=4$ the steps size was 1. }
\label{tab2}
\end{table}

%\subsection{Data Analysis}
The main part of the data analysis consists of finding the  minima of the free energy
and getting the Debye mass self consistent. To do the former we fit data sets for the free energy near its minima
%in the 2-dimensional space (of radius and holonomy). We fit this 
%set of data 
with a 2-dimensional parabola
\begin{eqnarray}
f &=& (v-v_0)M(v-v_0)+f_0
\end{eqnarray}
which has 6 variables. $v$ and $v_0$ are 2D vectors with $v$ containing the variables holonomy $\nu$ and radius $r$ and $v_0$ describing the correction to the point that were the minimum. $M$ is a 2 times 2 matrix with $M=M^T$ containing the coefficients for the fit.

The fit was done on the
 free energy values of $5^2=25$ points from a square, containing
5 points around the minimum.  The $6$ parameters from the fit are used as follows:(i)  $v_0$ and its uncertainties give the  values of densities and holonomy
at the minimum, plotted as results below;  (ii) the diagonal component of $M$ in
the holonomy direction was converted into the  value of the Debye mass $M_d$.
An additional requirement of the procedure, to make the ensemble approximately self-consistent,  is  that the Debye mass 
from the fit should be within
$\pm 0.25$ of the used input Debye mass value. 

To obtain the
 chiral properties -- such as the Dirac eigenvalue distributions and its  dependence on dyon number and volume --
we only used the ``dominant" configurations for each action S.

The results reported below, compare new results, for $Z_2$-symmetric QCD explained above,
to the ``old" ones, from II, for  
 $N_c=N_f=2$ QCD with antiperiodic fundamental  quarks.
%Our
%previous paper \cite{Larsen:2015tso} was devoted to the ``standard QCD":  in it
% both quark flavors are antiperiodic on $S^1$.
% Therefore both $u,d$ quarks interact with the $L$ dyons: this directly break  the $Z_2$ symmetry.
% The fermionic determinant squared suppresses the
%density of $L$ further, on top of suppression by the action: as a result 
%there is no really symmetric phase, as the densities are never equal, $n_L<n_M$.
%The confinement transition get weakened into a mere crossover, and so is the chiral transition.
%   
% 
% The ``center-symmetric QCD" is a new option studied in this paper. In it 
%  one quark flavor has a symmetric, and another  an antisymmetric
% periodicity.  
%Apart of such choice of the quark phases, the setting is exactly the same as 
%in the previous paper II: but as we will soon see, the results are drastically different.
 
 %  all with antiperiodic boundary conditions, it was seen that if one increased the amount of $L$ dyons, and thus increasing the amount of zero-modes, one would lower the free energy. This meant that the free energy preferred to have a lower value of the holonomy $\nu$. 

 \section{The holonomy potential and confinement}

% case the $Z_2$ symmetry is manifest, as   $L\leftrightarrow M$ dyon replacement. , the amount of zero-modes stay the same. The symmetry is apparent, since all $M$ and $L$ dyons are the same. This does not mean that every single configuration gives a free energy that has $Z_2$ symmetry, it means that the free energy does not change under $M \to L$, $L\to M$ and $\nu \to 1-\nu$.

%On simulations details , free energy measurements etc

%That this is indeed the case has been checked. For $n_L=n_M$ this means that the configuration should be symmetric around $\nu=\frac{1}{2}$, as is .

 \begin{figure}[h]
\includegraphics[scale=0.65]{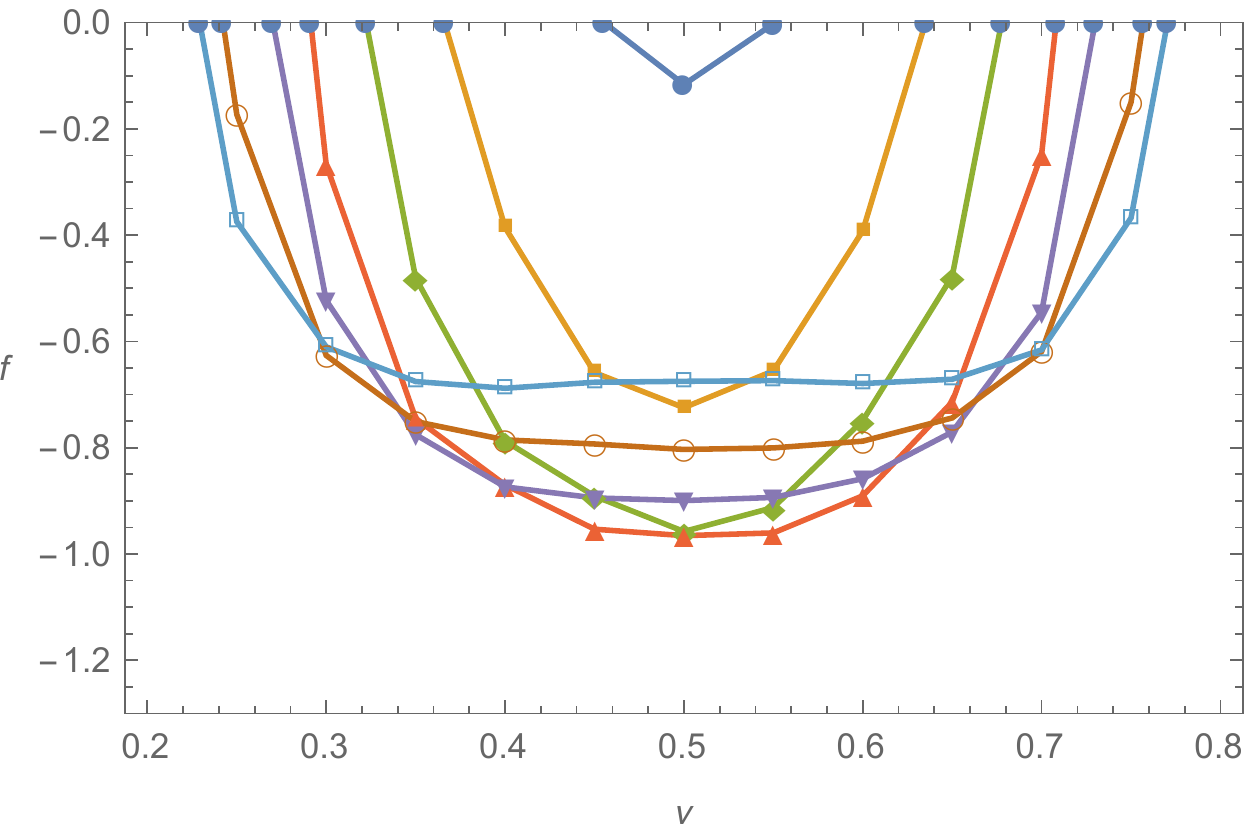}
\includegraphics[scale=0.65]{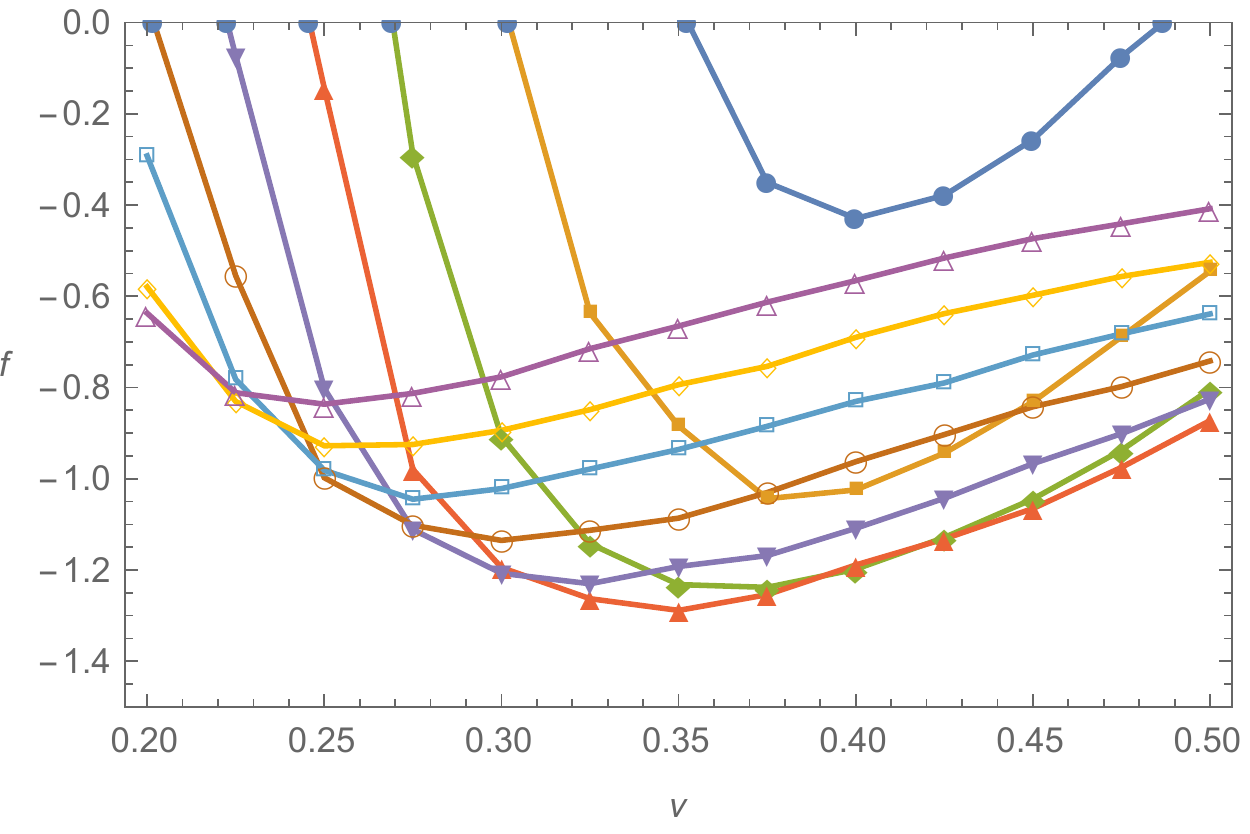}
\caption{
Free energy density as a function of the holonomy parameter $\nu$. The  upper plot is for $Z_2$-symmetric model and lower plot is for the model in which all quarks are
 anti-periodic. Different curves are for
different dyon densities. The densities are $(0.47,\bullet)$, $(0.37,\blacksquare)$, $(0.30,\blacklozenge)$, $(0.24,  \blacktriangle)$, $(0.20,\blacktriangledown)$, $(0.16,\circ)$, $(0.14,\Box)$, $(0.12,\diamondsuit)$, $(0.10, \vartriangle)$. Not all densities are shown. In both cases the action parameter is
$S=8.5$, and both dyon types are equally represented $n_M=n_L$. Note the dramatic
difference of the holonomy potentials for these two cases: the $Z_2$ potential is symmetric (for
equal dyon densities), while the periodic quarks produce an asymmetric minimum, and  thus slide smoothly
 towards smaller holonomies (to the left) as the dyon density decreases. 
}
\label{Md3_16}
\end{figure} 

The free energy resulting of the simulations are shown in Fig. \ref{Md3_16} as a function of holonomy value. Both for standard (lower plot) and $Z_2$-symmetric QCD (upper plot). 
At high density of the dyons
one finds a symmetric minimum for the $Z_2$-symmetric model. As the density decreases,
one finds behavior very different from both that of the quenched case (no quarks) with two minima
or in standard QCD, with broken center symmetry.

 While symmetry remains intact,  with the decreasing density (larger $S$) the minima
of the potential become very flat and wide. (A slight appearance of the minima  can be seen for the smallest density
which is not nearly as strong as in the quenched case.) We interpret this as an appearance of a large domain of
``mixed phase", a coexistence of many different configurations with different properties and $\nu$ but
degenerate free energy. 
The confining minimum in the middle is also found to be dominant for much larger range of densities.

\begin{figure}[t]
\includegraphics[scale=0.65]{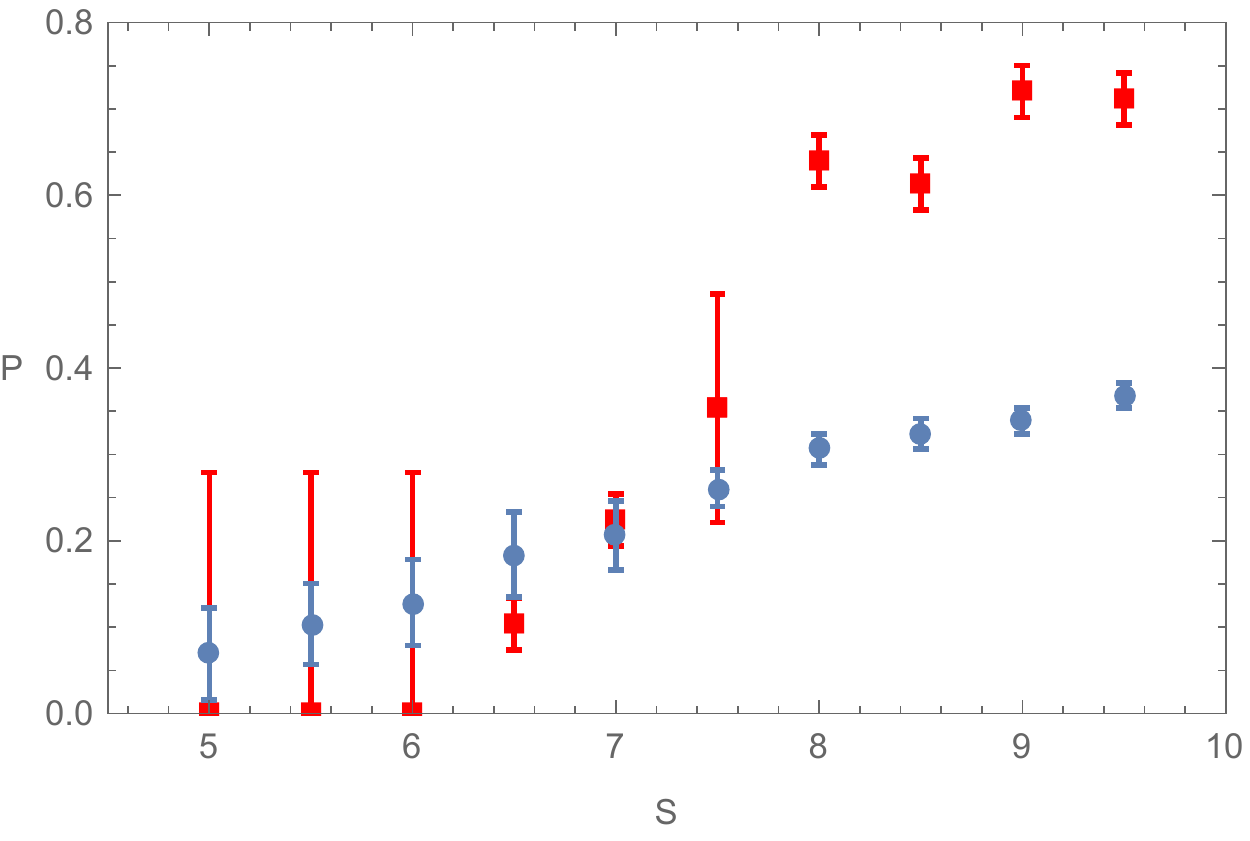}
\caption{ The mean Polyakov loop $P$ as a function of action parameter $S$, for $Z_2$-symmetric model
(red squares),
%%\label{Poly}
%(lower) 
%The mean Polyakov loop $P$ (blue) and the chiral condensate (red) for $L$ dyons $\Sigma$ as a function of action $S$, for 
compared to that for the $N_c=N_f=2$  QCD  with the usual anti-periodic quarks (blue circles). }
\label{Poly}
\end{figure}

Translating the location of the minimum to the mean Polyakov line, we plot the results in
Fig.  \ref{Poly}. It shows that while the two models under consideration have very similar
behavior at high densities of the dyons (smaller $S$ or the left side of the plot),    in the $Z_2$-symmetric 
model there appears a strong jump in $P$, from about 0.2 to 0.6.  Note that the intermediate point with large
error bar should be interpreted not as an uncertainty of the value, as the usual error bar, but rather as
reflection of the fact that in the ensemble the intermediate values of $P$ are all feasible, due to flatness
of the holonomy potential. In other words, this point is rather a vertical part of the curve,
indicative of strong first-order transition. This conclusion is consistent with lattice studies in \cite{Misumi:2015hfa},
in which the authors show some hysteresis curve for $P$, with a similar strong jump.

% and the holonomy only jumps away from the confining, when the simulation becomes difficult. In particular, the density of $L$
%dyons get too small, and the
%required Debye mass to maintain the configuration is hardly viable.

%The difference between two models can be also explained as follows.
%The quenched case behavior was ``unstable", in the following sense:
% increasing the amount of $M$ dyons relative to $L$ dyons,  the minimum got shifted towards smaller holonomy
% value  $\nu$, which enhances the effect. What happens in  the $Z_2$-symmetric case is  ``super-stability" or ``flattening": 
% increasing the fraction of $M$ dyons one observes very wide minima and flat minima, with 
%  almost the same free energy for 
% large range of the holonomy value, between $\nu=0.4$ to $0.6$.
% 
  
%What we therefore find is that at lower action/temperature, the dominating configuration lie around $\nu=0.4$ to $0.6$. Generally the dominating configuration is $\nu=0.5$, but it is very sensitive to small changes, and when the action becomes too small, below $S=5$, the factor $d_v$ that gives the density in the non-interacting gas, becomes so large, that configurations, slightly away from $\nu=0.5$ is dominating. 
%
%At higher action $S=7.5$ the holonomy jumps to around $\nu= 0.7$ or $0.3$. This jumps is observed in the debye mass, which goes from $2.5$ to $1.4$.

\begin{figure}[b]
\includegraphics[scale=0.65]{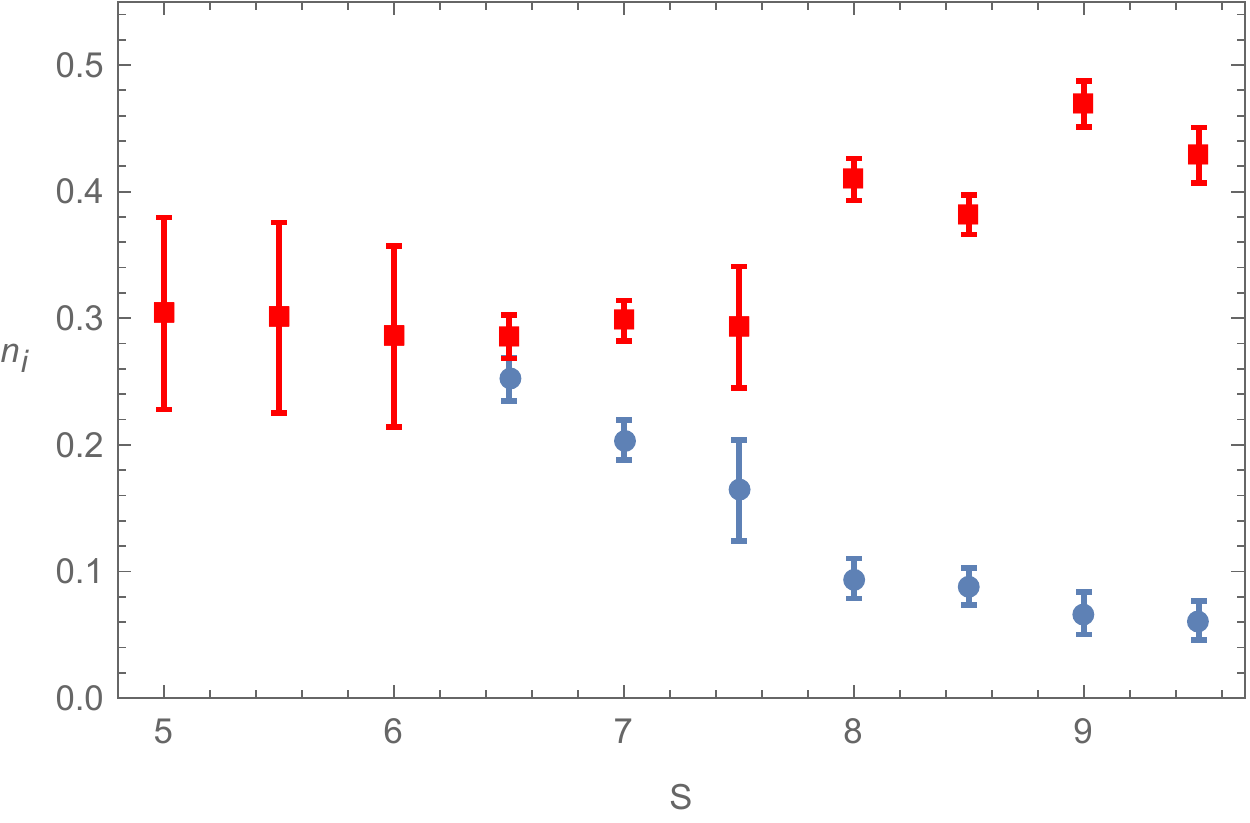}
\includegraphics[scale=0.65]{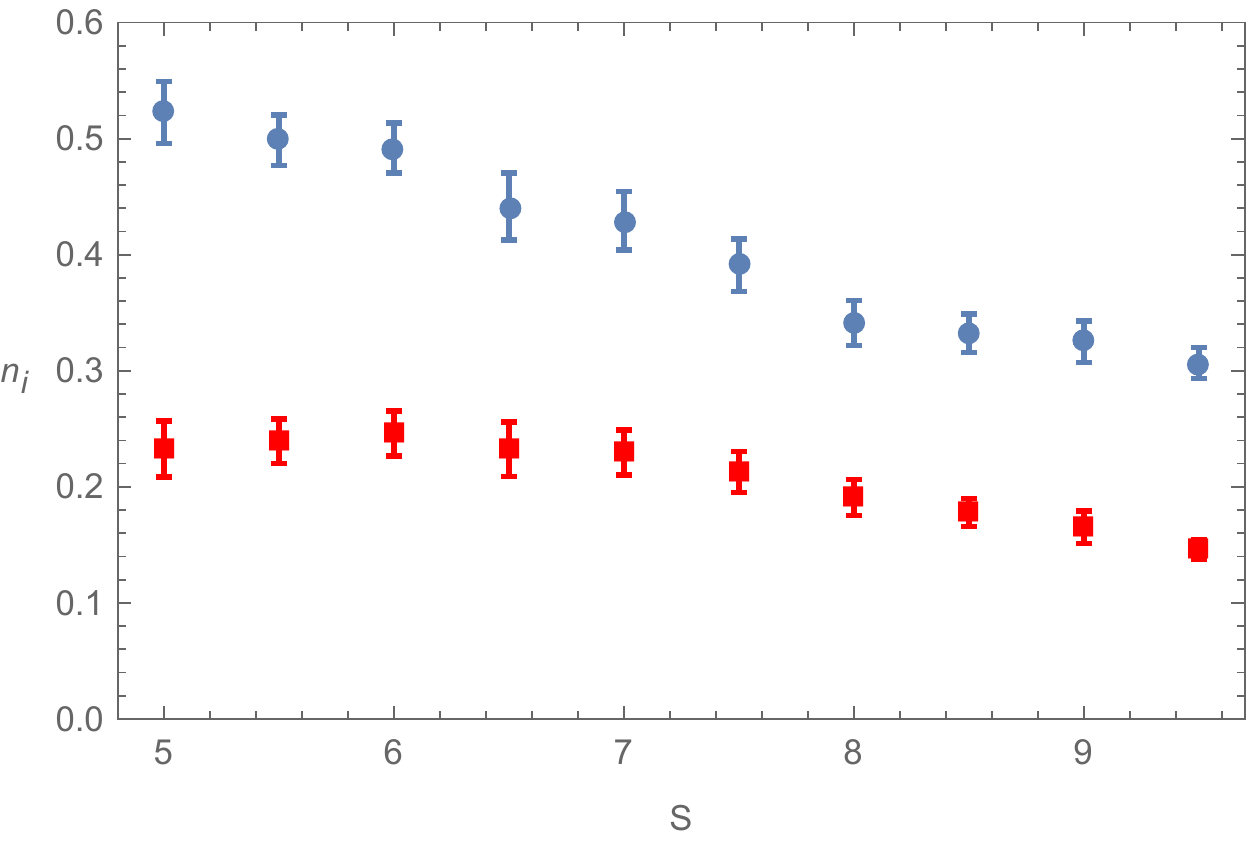}
\caption{ (upper) Densities of $L$ dyons (red squares) and $M$ dyons (blue circles), as a function of action parameter $S$,
for the $Z_2$-symmetric model. 
(lower) the same for the usual QCD-like model with $N_c=N_f=2$ and anti-periodic quarks.}
\label{Dense}
\end{figure}

The densities of the dyons in both models are shown in Fig. \ref{Dense}. The upper plot for the $Z_2$-symmetric model
display a very symmetric confining phase at the l.h.s. of the plot (small $S$, dense ensembles) complemented by
very asymmetric composition of the ensemble at  the r.h.s. The usual  QCD-like model with $N_c=N_f=2$  in the plot below shows that the $L-M$ symmetry never holds, due to only L-dyons coupling to the zero-modes, while the overall dependence on $S$ is much 
 less significant.

\begin{figure}[h]
\includegraphics[scale=0.65]{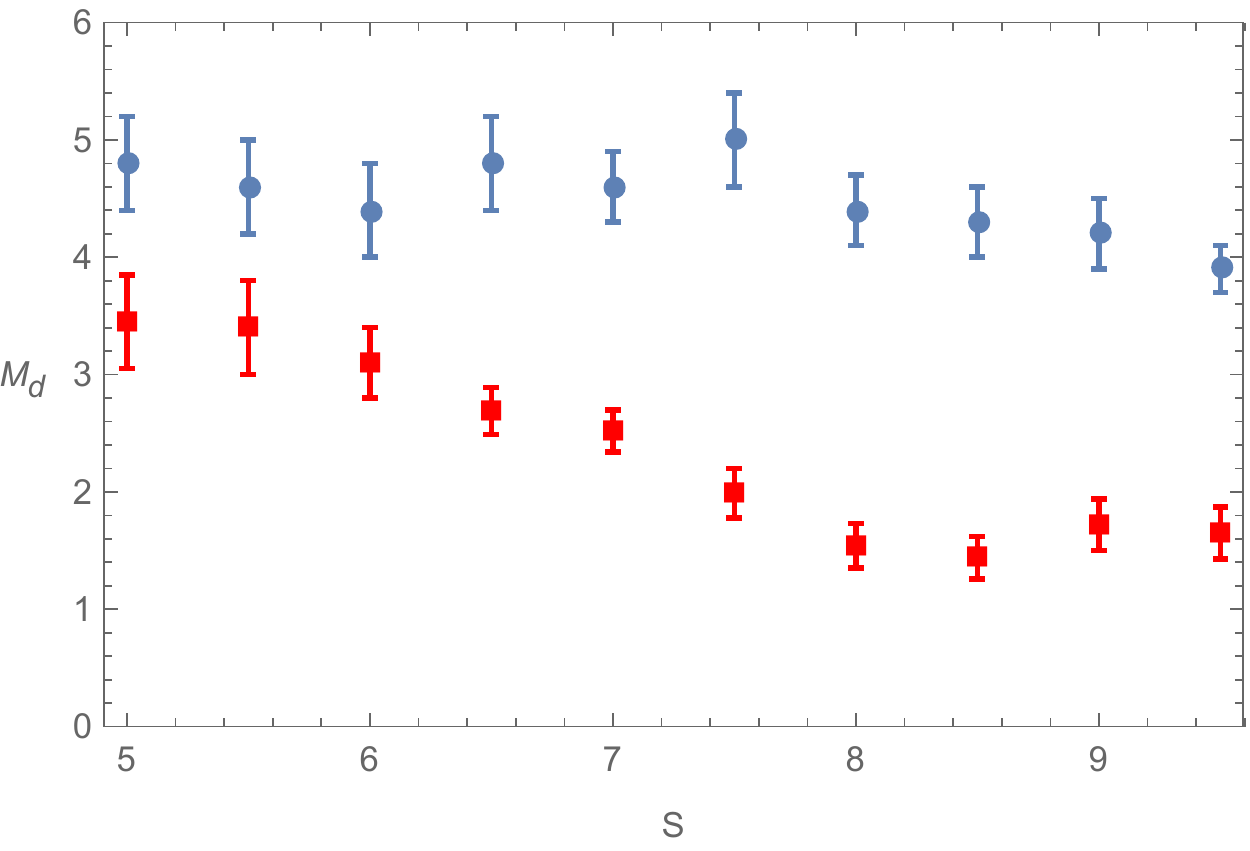}
\caption{Debye Mass $M_d$ as a function of action parameter $S$, for the $Z_2$-symmetric model
(red squares) and the usual QCD-like model with $N_c=N_f=2$ (blue circles).
}
\label{DM}
\end{figure}

Lastly, the Debye mass -- defined via the second derivatives of the effective potential at the minimum --
has been determined and plotted in
 Fig. \ref{DM}, again for both models. For the $Z_2$-symmetric model its values are 
 significantly lower than for the QCD-like model. Smaller mass indicate flatter potential and  stronger fluctuations,
 already discussed above.

%The inclusion of zero-modes on all dyons, means that we always have a symmetric situation between $\nu$ and $1-\nu$. For this reason we only show results for $\nu=0.5$ or smaller as shown in Fig. \ref{Poly}.
%
%Around the confining value of $\nu=0.5$ multiple configurations with very similar free energy exist. They have holonomy from $\nu = 0.415 $ to $0.585$. The symmetric configuration prefer the confining value $\nu=0.5$, while configurations with more $L$ dyons prefer smaller holonomy and configurations with more $M$ dyons prefer configurations with higher holonomy. A similar situations happened at $S=7$ where configuration between $\nu=0.43$ and $0.34$ showed very similar free energy. The points where several equivalent configurations exists, the average is shown, with big enough uncertainty to cover the area.
%
%Compared to results for the anti-periodic quarks Fig. \ref{PolyOld}, the $Z_2$ fermions Polyakov loop changes much more abrupt, especially around $S=7$ to $8$, where a quick transition from around $\nu=0.4$ to $\nu=0.3$ happens.
%
%The density, Fig. \ref{Dense}, is the same for both $M$ and $L$ dyons when the holonomy is $0.5$, but as the holonomy goes further and further away from the confining value, the densities become more and more different. Contrary to antiperiodic quarks Fig. \ref{DenseOld}, it is the $L$ dyons that become more abundant at higher temperatures.

\section{Chiral symmetry breaking}
As we already explained above, the main feature of the $Z_{N_c}$-symmetric model with $N_f=N_c$ distribute 
all types of quarks evenly, so that each type of dyons would have one quark flavor
possessing zero modes with it. This is in contrast to the usual QCD, in which all quarks are
antiperiodic and thus all have zero modes only with twisted $L$-type dyons.

The simplest examples considered in this work are two $N_c =N_f=2$ theories, the $Z_2$-symmetric model and the 
two color QCD.  In the former case the partition function
includes two independent fermionic determinants, one for $M$ and one for $L$
dyons, with a single quark species each. In the latter, one has a square (two-species) of the determinant
of hopping matrix over the $L$-dyons only. 

Here we remind well known facts about chiral symmetry breaking in such cases, and 
the consequences for such determinants. Theories with a single quark flavor have only
a single $U_a(1)$ symmetry,  broken explicitly by the fermionic effective action.
Indeed, it includes terms $\bar \psi_L \psi_R$ or $\bar \psi_R\psi_L$ directly coupling
components with opposite chiralities. So, there are no chiral symmetries to break,
and condensates are always nonzero, proportional to density of the topological objects.

The case with two or more flavors is different: there is the $SU(N_f)$ flavor symmetry, which can be
either broken or not, depending on the strength of the $2N_f$-quark effective interaction.

 \begin{figure}[t!]
\includegraphics[scale=0.65]{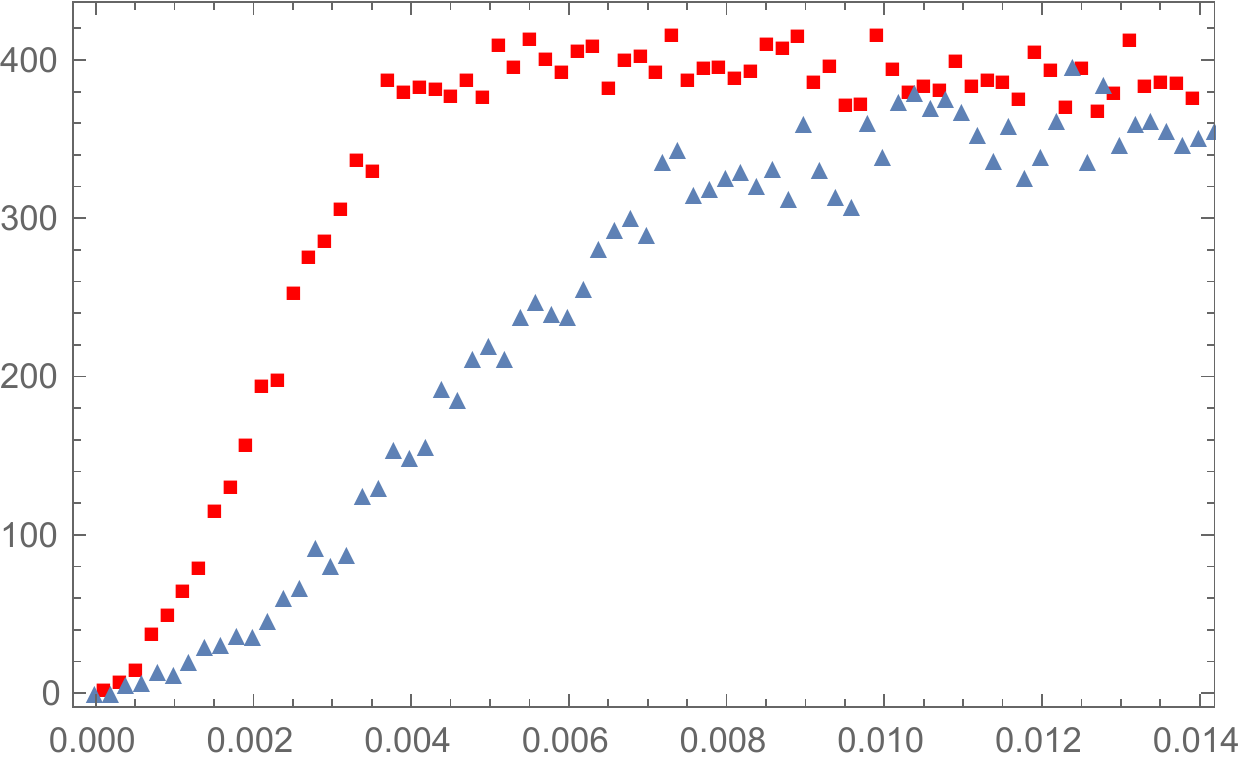}
\includegraphics[scale=0.65]{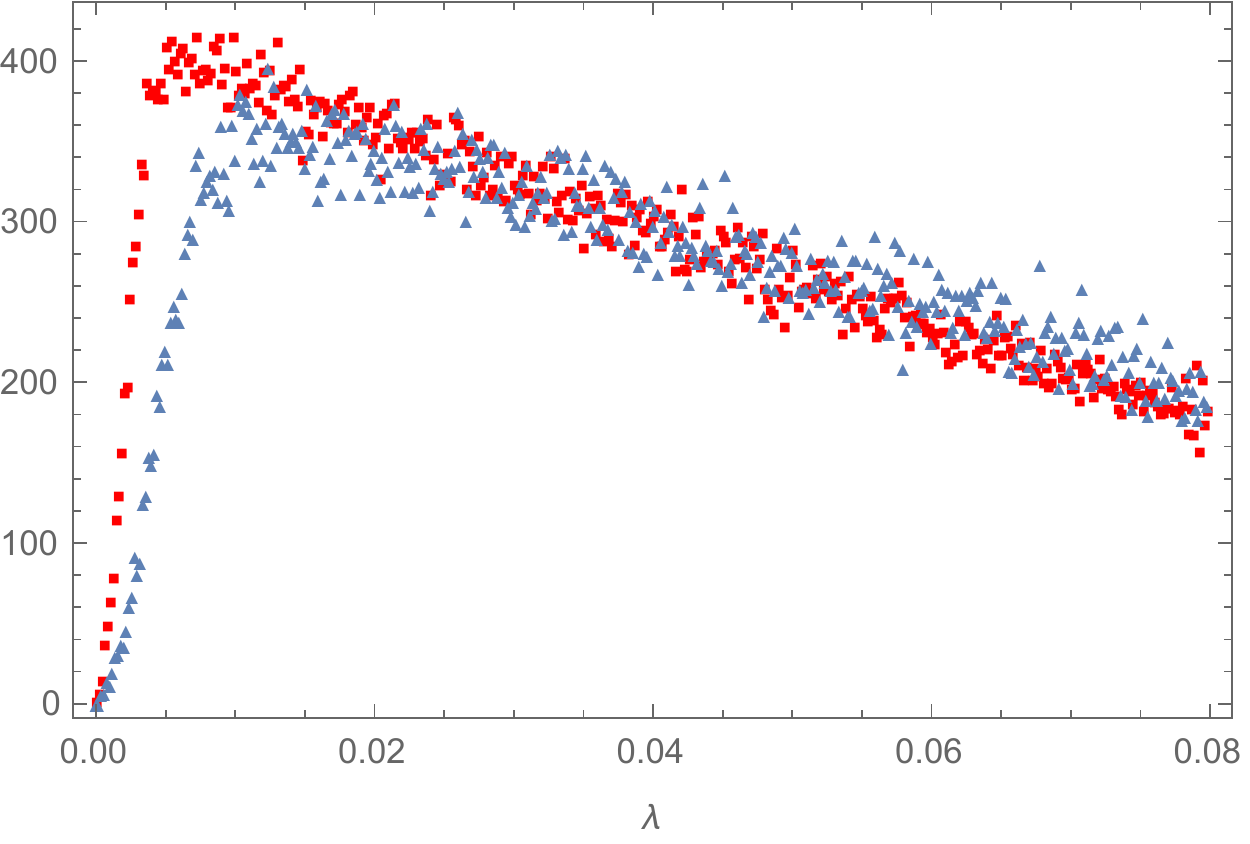}
\caption{The Dirac 
 eigenvalue distribution $\rho(\lambda)$ for ensembles of $64$ (Blue triangle) and $128$ (Red square) dyons,
 for $Z_2$-symmetric model at  $S=6$.
  The upper plot shows the region of smaller
  eigenvalues, in which one can see the finite volume ``dip", of a width which scales approximately as $1/V_4$ as expected.   The lower plot shows the same data sets, but in wider range
  of  eigenvalues: it displays the ``inverse cusp" shape of the distribution
 discussed in the text.
}
\label{Eigenvalues_S6}
\end{figure} 

\subsection{Dirac eigenvalue distribution}

Differences in chiral breaking mechanisms in these two models indicated above 
also manifest themselves in the Dirac eigenvalue distribution.

 For a proper perspective, let us remind that for the $SU(N_f)$ flavors with $N_f \geq 2$ a general Stern-Smilga theorem \cite{Smilga:1993in}
states that the eigenvalue distribution at small $\lambda$ has the so called ``cusp" singularity

\be \rho(\lambda) = \frac{\Sigma}{\pi} \left(1+\frac{| \lambda | \Sigma (N_f^2-4)}{32 \pi N_f F_c^4}  +...  \right) 
\label{smilgastern}\ee

For $N_f>2$ the coefficient is positive -- this is known as ``direct cusp", and was also observed, both
on the lattice and in the instanton models. 
In the particular case $N_f=2$ this cusp is absent: this fact can be traced to the absence of
symmetric $d^{abc}$ structure constant in the case of $SU(2)$ group. 
Indeed, both the calculations done in the instanton liquid framework (for examples and references see \cite{Schafer:1996wv}) and our previous studies II of the $N_f=2$ theory had produced ``flat" eigenvalue distribution

\be 
 \rho_{N_f=2}(\lambda)\sim const \ee

In the $N_f=1$ case the distribution does have a singularity at $\lambda=0$ of the form of the ``inverse cusp", $\sim -| \lambda |$, with 
$negative$ coefficient. The Stern-Smilga derivation does not apply, but the theorem has been rederived for general $N_f$ using partially quenched chiral perturbation theory in \cite{Osborn:1998qb}.%, but empirically it has been observed that
%the distribution does have a singularity at $\lambda=0$ of the form of the ``inverse cusp", $\sim -| \lambda |$, with 
%$negative$ coefficient. %We therefore look

 Our results for  the $Z(N_c)$-QCD under consideration shown in Fig. \ref{Eigenvalues_S6}
also show  the ``inverse cusp" with linear behavior of $\rho(\lambda)$. (We use this fact to
extrapolate  $\rho(\lambda)$ to $\lambda\rightarrow 0$ and to extract the value of the quark condensate and the value of the coupling constant $F_c$.)
In the other model, the $N_c=N_f=2$ QCD, such ``inverse cusp" is absent, see II.

So far our discussion assumed an infinite volume limit, in which case the  Dirac eigenvalue spectrum extends 
till $\lambda=0$. However, it is well known that any
finite-size systems, with 4-volume $V_4$,  have  the smallest
 eigenvalues of the order  $O(1/V_4)$. This creates  the so called ``finite size dip", in the 
 eigenvalue distribution, also clearly visible in  Fig. \ref{Eigenvalues_S6}(upper).
 One can see that doubling of the volume, from 64 to 128 dyons at the same density,
 reduces the width of this dip roughly by a factor two, as expected.
 
% Its exact shape can be been studied in the chiral  random matrix theory framework \cite{},
% to which our numerical distributions in II has been compared. For definiteness, we mention its shape
%% 
%%In the case of two fundamental fermions, it was found in ???, that the eigenvalue distribution around $\lambda =0$ could be fitted reasonably well with the distribution for $SU(2)$ for $N_f=2$ taken from ???. 
%for $N_f=2$ and massless quarks
%
%\fbox{to be changed to Nf=1}
%
%\begin{eqnarray} \label{CCfit}
%f(x ) &=& \frac{x}{2}(J_2(x)^2-J_1(x)J_3(x))+\frac{1}{2} J_2(x)(1- \int _0 ^x J_2(t) dt)
%\end{eqnarray} 

%\begin{figure}[h]
%\includegraphics[scale=0.65]{Eigenvalues_Nm16_Md3_nu05_r14.pdf}
%\caption{$N_f=16$, $r=1.4$ $\nu=0.5$ $M_d=3$.}
%\label{Eigenvalues_Nm16_Md3_nu05_r14}
%\end{figure} 

%It is observed that the eigenvalue distribution at the top is not completely constant but fall with a small slope. This explain part of the offset between the distributions at $N=64$ and $N=128$, but doesn't explain the full difference. In the same way the scaling ration between the two curves also tend to be about $10\%$ too large.

\begin{figure}[h]
\includegraphics[scale=0.65]{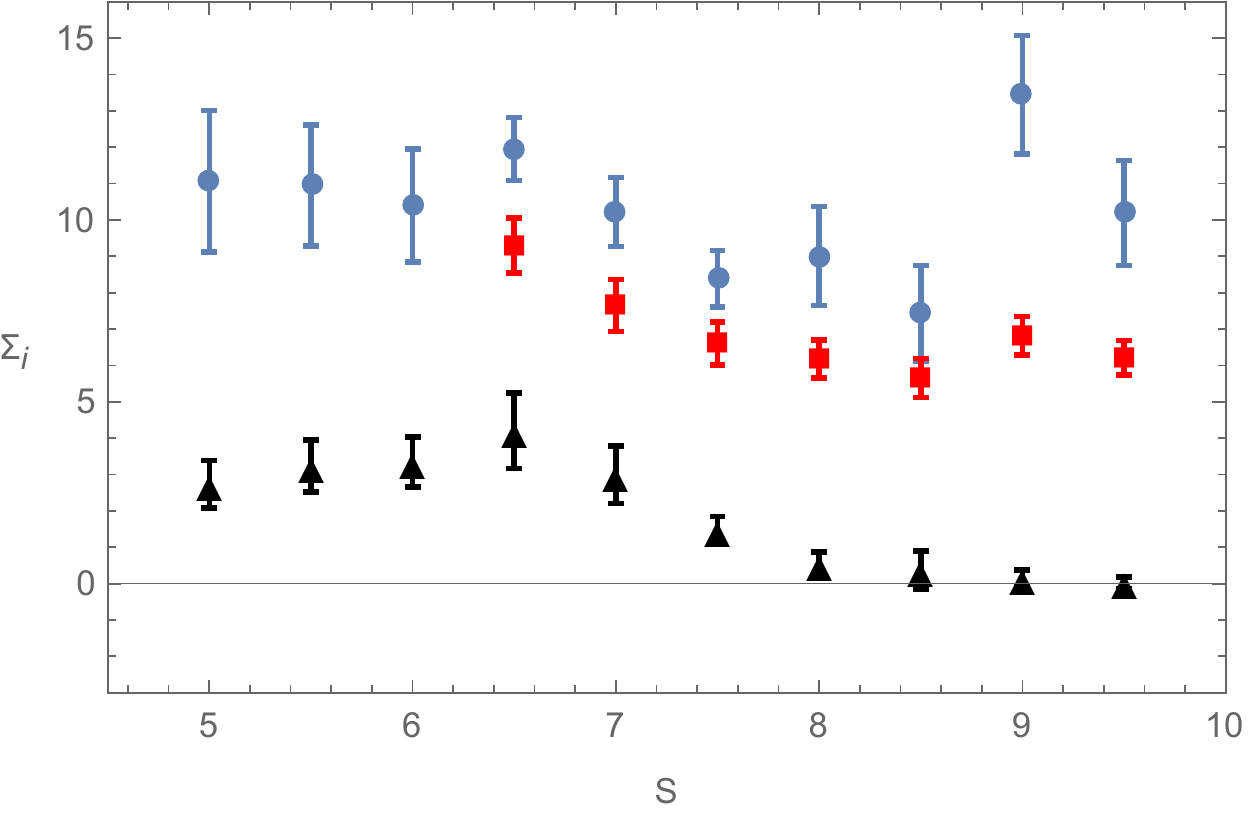}
\caption{Chiral condensate generated by $u$ quarks and $L$ dyons (red squares) and $d$ quarks
interacting with $M$ dyons (blue circles) as a function of action $S$, for the $Z_2$-symmetric model.
For comparison we also show the results from II for the usual QCD-like model with $N_c=N_f=2$
by black triangles.  }
\label{CC}
\end{figure}

As the holonomy jumps away from its confining value $0.5$, the dyon densities  become different. Unlike the fundamental quarks, where the holonomy goes down, the densities of $L$ dyons become larger than that of $M$ dyons. The total density goes down, but the reduction in $M$ dyons leaves space for a few more $L$ dyons. This means that on one hand the density is larger for $L$ dyons, and the zero-mode density is therefore higher. On the other hand, the factor in the exponential in $T_{ij}$ (Eq. \ref{Tij}) is $\bar{\nu}=1-\nu$ for $L$ dyons, and $\nu$ for $M$ dyons. This means that as $\nu$ becomes smaller, the effective density of the zero-modes associated with $L$ dyons become smaller, while the zero-modes associated with $M$ dyons gets an increased effective density. It is therefore the interplay between these two effects, that control which of the condensates are largest.
This results in what we show in Fig. \ref{CC}, where the $M$ dyon condensate appears to be slightly larger than the $L$ dyon condensate, and both condensates decreases slightly in accordance with the total density of dyons.
It is also observed that each gas of zero-modes effectively works as a $N_f=1$ ensemble, with
   non-vanishing condensates even at the lowest densities we studied \footnote{It should be noted that the chiral condensate is harder to study, as the amount of dyons in the simulation becomes small, which happens for $M$ dyons when there is a large asymmetry in the density. This explains larger error bars for one of the condensate and 
   why we stopped our studies at those particular parameters.}
    (the r.h.s. of the plot).
   The other model -- $N_c=N_f=2$ QCD --has a condensate shown by black triangles: it clearly has 
   chiral symmetry restoration. At $S>8$ we detected no presence of a condensate.

%\includegraphics[scale=0.5]{Fit_Md_square.jpg}
%\caption{Debye Mass $M_d$ as a function of action $S$ for the usual QCD-like model with $N_c=N_f=2$ and anti-periodic quarks.}
%\label{DMOld}
%\end{figure} 

%It is important to note that the configurations are less stable in the case of $Z_2$-QCD. The region around the confining value to the area where the holonomy is about $0.3$ ($P=0.6$) is only seen when using minimums in both radius and holonomy. In case the minimization of the free energy is only done in holonomy space, then one observes a jump from the grey area and up to around $P=0.6$, where the jumps happens around $S=7.5$. It can therefore be argued that that the area of $P=0$ to $0.5$ is part of the hysterical area, and only around $P=0.6$ has a definitive new minimum emerged. 

\begin{figure}[h]
\includegraphics[scale=0.65]{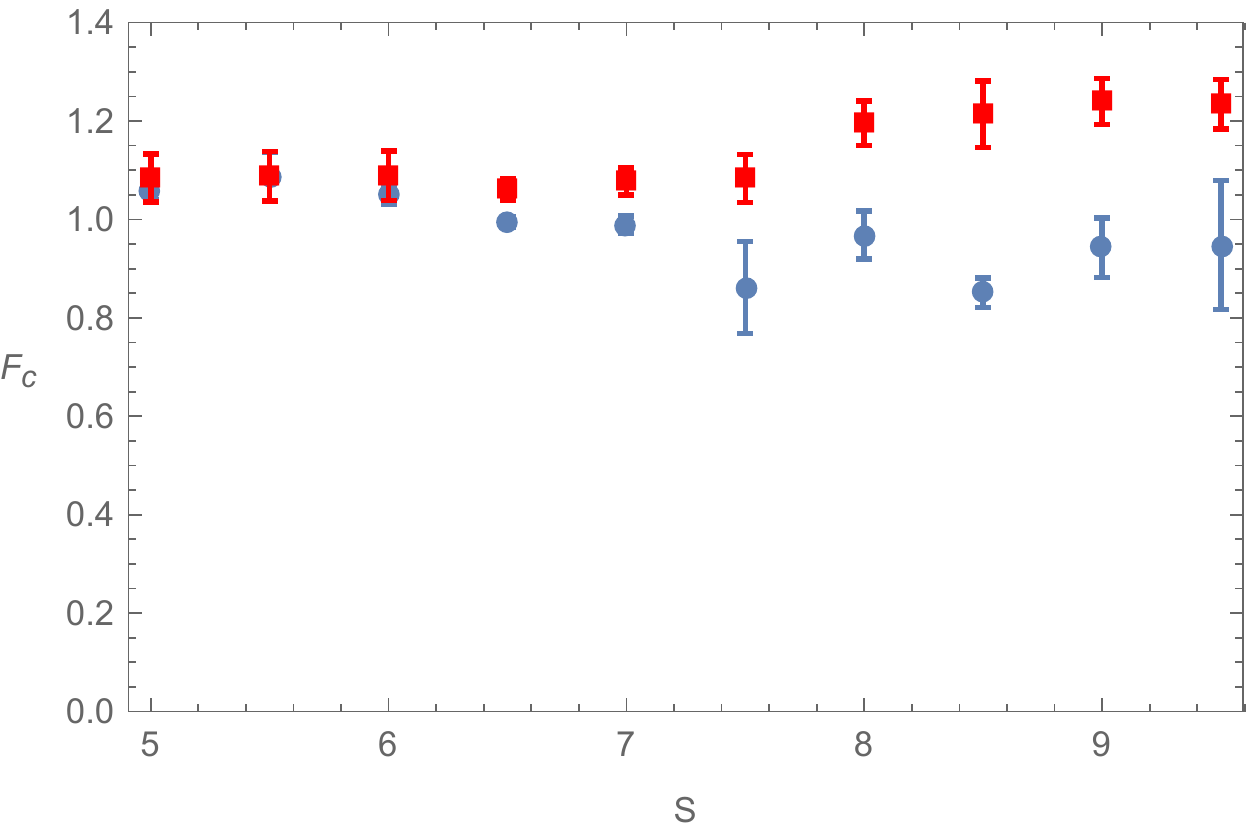}
\caption{Coupling constant $F_{c}$ for the $M$ dyon ensemble (blue circle) and the $L$ dyon ensemble (red square). }
\label{F}
\end{figure}

The coupling constant $F_c$ (Fig. \ref{F}), obtained from the slope of the eigenvalue distribution and 
Smilga-Stern theorem (\ref{smilgastern}),
is nearly density-independent:
%Part of this is due to how $F_c$ is obtained from the slope $F_c\sim (slope)^{-1/4}$, but it is also due to a slowly changing slope, that 
it changes by a factor of around $1.5$ from $S=5$ to $S=9.5$. This is consistent with behaviour of the quark condensate, and similarly indicate that in the $Z_2$ model the chiral symmetry does not show tendency to be restored.

\section{Summary and discussion} 
To put it in perspective, let us start by briefly reminding the main finding of the previous two papers of these series.
In I we had shown that in pure gauge theory with the $SU(2)$ color the instanton-dyon ensemble
undergoes confinement transition as the dyon density reaches certain critical value. The 
high-density confining phase has holonomy $\nu=1/2$ and equal densities and other properties of 
all types of instanton-dyons. 

In paper II we added two light fundamental antiperiodic quarks, as they are in the $SU(2)_c, SU(2)_f$
QCD. Deconfinement transition gets significantly smoothed to a crossover. Chiral symmetry 
transition is also somewhat smooth, and happens when the Polyakov loop gets close to the confining value $P=0$. Thus we concluded there,
that the old question of interrelation of the two transitions is finally over: large enough  instanton-dyon density
does both.

In this work we introduce flavor holonomies, and, following  Kouno et al \cite{K1}  arrange them into $Z_2$-symmetric model,
which we compare to theories with all-periodic
or all antiperiodic (QCD) quarks. The results are dramatically different: the $Z_2$-symmetric model has 
very symmetric confining phase and a quick
 deconfinement transition, but no apparent chiral symmetry 
restoration!
The deconfinement becomes much stronger, a first order transition with clear mixed-phase for intermediate
dyon densities.  Chiral symmetry seem to be never restored, even for the smallest densities, as indeed is expected
based on analogy to the one-flavor QCD. Different flavors do have different condensates: but the
difference in condensates is much smaller than the difference in the dyon densities.
Our  approach, based on instanton-dyons, provides the simplest explanation of these observations. Flavor-dependent periodicity condition effectively manipulate their coupling to dyons of different kinds.   
The  $Z_{N}$-symmetric model is the ``most democratic"arrangement, producing
basically $N$ copies of single-flavor topological
ensembles,  drastically different from one $N$-flavor ensemble of $L$-dyons
in the usual QCD.  

So, we take lattice confirmation of these phenomena,  by Misumi et al  
\cite{Misumi:2015hfa}, as basically a confirmation of its main statement:   chiral symmetry breaking is
induced by  zero modes of  the instanton-dyons. Needless to say, much more detailed studies on the lattice
are possible: perhaps direct identification of the quasi-zero mode localization with  the instanton-dyons
in the gauge field ensemble would soon be possible.
 
%Those were referred to as  ``most democratic" and  ``most exclusive" arrangements
%above. We think that this good agreement is the most clear manifestation of the large role
%of the instanton-dyons in forming deconfinement and chiral phase transitions.

 {\bf Acknowledgements.}
We would like to thank T.~Iritani and I. Zahed for useful discussions. This work was supported in part by the U.S. D.O.E. Office of Science,  under Contract No. DE-FG-88ER40388.

%%%%%%%%%%%%%%%%%%%%%%%%%%%%%%%%%%%%%%%%%%%%%%%%%%%

\end{document}